\documentclass[aps,pra,twocolumn,superscriptaddress,nobibnotes,letterpaper]{revtex4-1}
\usepackage[pdftex]{graphicx}

\usepackage{graphicx}
\usepackage{amsmath}
\usepackage{verbatim}
\usepackage{color}

\usepackage[colorlinks=true, allcolors=blue]{hyperref} 

\begin{document}

\title{Modeling Green's functions measurements with two-tip  scanning tunneling microscopy}

\author{Maarten Leeuwenhoek}
\affiliation{Kavli Institute of Nanoscience, Delft University of Technology, Lorentzweg 1, 2628CJ Delft, The Netherlands}
\affiliation{Leiden Institute of Physics, Leiden University, Niels Bohrweg 2, 2333CA Leiden, The Netherlands}

\author{Simon Gr\"{o}blacher}
\email[]{s.groeblacher@tudelft.nl}
\affiliation{Kavli Institute of Nanoscience, Delft University of Technology, Lorentzweg 1, 2628CJ Delft, The Netherlands}

\author{Milan P. Allan}
\email[]{allan@physics.leidenuniv.nl}
\affiliation{Leiden Institute of Physics, Leiden University, Niels Bohrweg 2, 2333CA Leiden, The Netherlands}

\author{Yaroslav M. Blanter}
\email[]{y.m.blanter@tudelft.nl}
\affiliation{Kavli Institute of Nanoscience, Delft University of Technology, Lorentzweg 1, 2628CJ Delft, The Netherlands}

\begin{abstract}
A double-tip scanning tunneling microscope with nanometer scale tip separation has the ability to access the single electron Green's function in real and momentum space based on second order tunneling processes. Experimental realization of such measurements has been limited to quasi-one-dimensional systems due to the extremely small signal size. Here we propose an alternative  approach to obtain such information by exploiting the current-current correlations from the individual tips, and present a theoretical formalism to describe it. To assess the feasibility of our approach we make a numerical estimate for a $\sim$25~nm Pb nanoisland and show that the wavefunction in fact extends from tip-to-tip and the signal depends less strongly on increased tip separation in the diffusive regime than the one in alternative approaches relying on tip-to-tip conductance.
\end{abstract}

\maketitle

\section{Introduction}

Green's functions provide a general framework for perturbed and interacting electron systems. Direct experimental access to the electron Green's function is vital for our understanding of (new) complex systems and electronic states of matter. This access is increasingly provided by the emergence of spectroscopic imaging STM (SI-STM) and by various types of angle resolved photo emission (ARPES) experiments.
A further possible tool for probing the single electron Green's function locally is double-tip STM, where two tips are brought into tunneling simultaneously within few (tens of) nanometers apart~\cite{Niu1995,byers1995probing,SettnesPRL}. The challenge of accessing the Green's function using a two-probe setup is twofold: (i) Since it is a second order tunneling process, the signal depends strongly on both the tip-to-sample and tip-to-tip distance~\cite{Niu1995,byers1995probing,settnesPRB}, and (ii) experimental realization of such a small tip separation in combination with the stringent stability requirements STM brings has proven challenging and has been a long standing goal for the multiprobe community~\cite{Hasegawa}. 

Here we explore and present a theoretical formalism for an alternative approach which, using a double-tip STM, has access to the propagator --- an averaged product of two single-electron Green's functions. The propagator determines the nature of electron wave propagation and is essential for understanding quantum effects in electron transport. Here, we show that it can be measured locally, between the points corresponding to the tip positions. The advantage of this approach is that, compared with the approach mentioned above, the result is not exponential in terms of tip-to-tip distance, at the cost of having a higher power of the tip-sample tunneling amplitude. We concentrate on the diffusive regime of electron transport and show that by calculating statistical correlations between the individual currents from the tips to the sample we can access the diffusion propagator at the nanoscale. Using the proposed formalism, we performed initial numerical estimates on Pb nanoislands to demonstrate the feasibility of this approach.     

Much progress has been made over the past decades towards a stable, well-controlled double-tip microscope~\cite{voigtlander2018invited} able to probe the local Green's function by reducing tip radii and increasing their aspect ratio~\cite{Kolmer2017,kubo2006epitaxially,konishi2007high,yoshimoto2007four}, low temperature and/or ultra high vacuum operation~\cite{shiraki2001independently,kim2007cryogenic,hobara2007variable}, mechanical stability~\cite{guise2005development,ma2017upgrade}, and navigation of the tips~\cite{okamoto2001ultrahigh,Kolmer2017}. Recently we have seen a re-emergence of the double-tip STM~\cite{Kolmer2017,thamankar2013low} culminating in the first two-point single electron Green's function measurements to date using a multiprobe system on quasi-one-dimensional dimer rows on the Ge(001) surface~\cite{kolmer2019electronic}.  

In parallel we observe a similar resurgence of nanofabricated STM probes~\cite{siahaan2015cleaved,ciftci2019polymer,leeuwenhoek2019nanofabricated} that can be equipped with two (fixed) probes that are compatible with ultra high vacuum and low temperature operation and potentially allow the integration in ultra-stable single tip STM systems currently available~\cite{leeuwenhoek2019nanofabricated}. Advances in modern nanofrabrication techniques such as focussed ion beam milling or electron beam induced deposition could lead to a tip separation of a few tens of nanometers in the very near future. Driven by this experimental progress we outline the theory for our new measurement formalism here and make a numerical estimate to assess its feasibility.

In Sec.~\ref{sec:II} we recap earlier proposals for measuring the electron Green's function with a double-tip STM. Subsequently, in Sec.~\ref{sec:III} we discuss the current correlations and show that they are proportional to the diffusion propagator. In the same section, we produce numerical estimates of the effect and show that is can be measured using the current technology. We present conclusions in Sec.~\ref{sec:IV}. Some technical details from the derivation of Sec.~\ref{sec:III} are relegated to the Appendix.

\section{Measuring Green's functions with STM involving two tips} \label{sec:II}

\begin{figure}[t]
	\begin{center}
		\includegraphics[width=1\columnwidth]{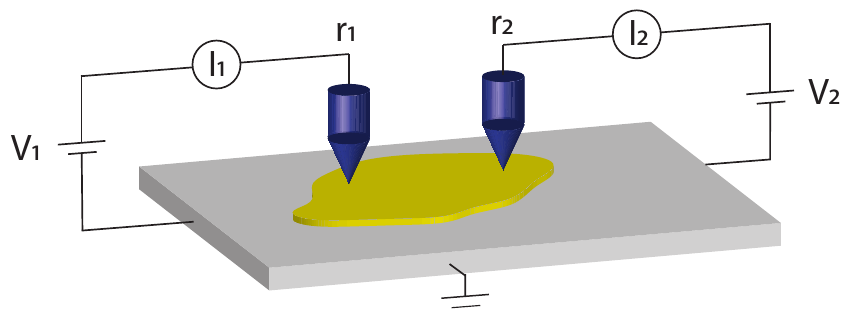}
		\caption{Schematic of a two-tip setup on a mesoscopic island.}
		\label{fig:DTschematic}
	\end{center} 
\end{figure}

Before we outline our alternative tools that use current-current correlations to probe the electronic states we first briefly introduce the method proposed by Refs.~\cite{Niu1995, byers1995probing} that shares the same three-terminal setup. The tips and the sample are kept at constant chemical potentials: Tip 1 ($\mu_1$), tip 2 ($\mu_2$), and the sample ($\mu_0$) and, similar to single probe STM, both tips are individually biased ($V_1$, $V_2$) and their respective currents measured ($I_1$, $I_2$) as shown in Fig.~\ref{fig:DTschematic}. The response of currents to voltages is described by the conductance matrix $\sigma_{ij}$, $I_i = \sigma_{ij}V_j$.


In a usual STM experiment the only response of the current in a tip to the voltage applied beween the tip and the sample is measured. For the two-probe setup this translates into diagonal elements of the conductance matrix, $\sigma_{11}$ and $\sigma_{22}$, which are proportional to the local density of states (LDOS) at the locations of the tips. With the two-probe setup one can also obtain the off-diagonal elements $\sigma_{12}$ and $\sigma_{21}$ that contain information about transport properties of the electrons inserted at one tip and collected on the other. In fact, this complementary information should allow -- in principle -- to obtain the full single electron Green's function~\cite{Niu1995}.  

The transport is described by a co-tunneling process with the sample as intermediate state. The properties of this process can be derived using Fermi's golden rule to second order and result in a transconductance $\sigma_{21} = \partial I_2/\partial V_1$~\cite{chan1997theories}. The same expression, albeit in slightly different form was obtained in the original work~\cite{Niu1995,byers1995probing} when looking at second order transport,
\begin{equation}
\sigma_{21}=\frac{\partial I_2}{\partial V_1}=\Gamma_1\Gamma_2\frac{2\pi e^2}{\hbar} |G(\mathbf{r}_1,\mathbf{r}_2;\epsilon=\mu_1)|^2,
\label{eq:Sigma21}
\end{equation}
where $G(\mathbf{r}_1,\mathbf{r}_2;\epsilon)$ is the retarded Green's functions of the sample for non-interacting electrons at zero temperature. We note that the signal size is now quadratic in tip-sample coupling and $|G(\mathbf{r}_1, \mathbf{r}_2)|$ is on the order of $10^{-2}$ for 2D systems and a tip-to-tip distance of a few tens of nanometers and is inversely proportional to that distance~\cite{Niu1995}. A similar result can be obtained using the Landauer formalism following Settnes \textit{et al.}~\cite{SettnesPRL,settnesPRB}. Unsurprisingly, the first double-tip STM results were taken on a quasi-one-dimensional system~\cite{kolmer2019electronic} where the signal is stronger overall and does not decay with tip separation.  



\subsection{Technical considerations}

The technical development of the double-tip STM's has been led by the multi-probe community that originally focused on studying resistance in mesoscopic systems on (sub)micrometer length scales by contacting the surface with -- ideally -- four probes. An intrinsically more simple double-tip STM designed for Green's function mapping uses only two tips to probe the single electron Green's function; however it requires operation in the tunneling (not contact) regime, prolonged out-of-feedback measurements and nanometer tip separation. Therefore mechanical stability and tip-to-tip distance make up the main challenges that need addressing.

For the latter, the radii of curvature of the tips need to reduced to the (tens of) nanometer range to achieve tip separation that is effectively set by twice the tip radius. Tips made by well-controlled tungsten etching, sharpened by FIB milling or equipped with metallized carbon nanotubes are used to create extremely sharp, high aspect ratio tips capable of achieving tip separations down to 30~nm~\cite{Kolmer2017}. Being able to navigate the two tips to such proximity that the scanning range of each tip overlaps has proven challenging as well, especially without any optical input. Several solutions have been explored~\cite{okamoto2001ultrahigh,tsukamoto1991twin} resulting in an additional scanning electron microscope (SEM) column as the most common solution~\cite{Jaschinsky2006,yang2016imaging,cherepanov2012ultra}. The separate piezo drives for each tip together with the SEM column makes for a more complex and elaborate apparatus, making it more challenging to achieve very low temperatures and rival the stability of compact single tip STM designs~\cite{voigtlander2018invited,ma2017upgrade}, but recently that is changing~\cite{yang2016imaging} and has paved the road to the first transconductance measurements to date~\cite{kolmer2019electronic}.

Some of the navigation and stability issues can also be overcome by relocating complexity of two probes from the STM head design to the tip itself, {\em i.e.} by having two nanofabricated tips fixed on a single device~\cite{leeuwenhoek2019nanofabricated,gurevich2000a,bo2000scanning,nagase2008}. Such devices can indeed be implemented in (commercially available) ultra stable single tip STM systems that operate at low temperature and UHV conditions~\cite{leeuwenhoek2019nanofabricated}. Future application for controlled double-tip experiments will depend on the ability to get both tips into tunneling simultaneously with good stability.

\maketitle

\section{Propagator from the current correlations} \label{sec:III}

\subsection{Theoretical consideration for measuring the diffusion propagator}

To further assist the recent experimental progress we outline the theory for a novel, alternative, experiment using two tips on a nanoisland for probing the propagator $\Pi(\mathbf{r_1},\mathbf{r_2})$ at the nanoscale and highlight the feasibility of our approach with numerical estimates. Whereas we specialize on and make estimates for the regime of diffusive motion of electrons, the principle is more general and applies to any underlying electron dynamics. We note that this is just one example of accessing Green's functions with double-tips, others are mentioned above~\cite{Niu1995,byers1995probing,SettnesPRL,settnesPRB,Ruitenbeek2011,Buttiker1998,Buttiker1999}. Our example has the advantage of being simpler and yielding an improvement of the signal strength. It is based on correlating the two measured single tip currents in order to obtain correlations between electron states at the respective tip positions and to ultimately measure the electronic diffusion propagator on the nanoscale.

In this Section, we first describe the specific setup we are considering here. We then derive an explicit expression of the diffusion propagator as a function of our experimental observables. By exploiting the formalism of level and wavefunction statistics developed earlier~\cite{Blanter1997,Mirlin2000}, we show that the correlations of the amplitude of the same wave function are sustained even at these relatively large distances (much larger than the Fermi wavelength), while the correlations between different wavefunctions decay, resulting in the expression that relates the measured currents to the diffusion propagator. Finally, we apply the formalism to metallic nanoislands and provide some numerical estimates that yield the required sizes of the double-tips.

\subsubsection{Preliminaries}

The scenario considered here consists of two tips held at individual bias voltages $V_i$ while measuring the individual currents $I_i$. The sample is grounded, which considerably simplifies the experiment. We assume a smart tip consisting of two tips located at $\mathbf{r_1}$ and $\mathbf{r_2}$, with the distance between the tips being much longer than the Fermi wavelength. In addition, the tip-to-sample coupling is the same for both tips. As a starting point for the theoretical description of the tunneling process we consider the Tersoff-Hamann model of STM~\cite{chen1993introduction}, with the tunneling current from each tip to the substrate being~\cite{Lounis} 
\begin{equation} \label{Tersoff}
I (\mathbf{r}, V) = A \int_0^{eV} n^s (\mathbf{r}, E) dE \ .
\end{equation}
Here, $E$ is the energy and $A$ is the tip-sample coupling which includes details of the tunneling process, of the tip, and is exponentially dependent on the tip-to-sample distance. The order of magnitude of $A$ is $G_T/\nu$, with $G_T$ and $\nu$ being the tunneling conductance (tip to substrate) and the density of states (per volume) in the substrate. The information about the substrate is encoded in the function $n^s$,
\begin{equation} \label{ns}
n^s (\mathbf{r}, E) = \sum_k \delta \left(E - E_k \right) \left\vert \psi_k (\mathbf{r}) \right\vert^2 ,
\end{equation}
where $E_k$ and $\psi_k$ are the exact eigenvalues and eigenfunctions of an electron in the substrate. 

In the following, we focus on weakly disordered metals, where $k_Fl \gg 1$, $k_F$ and $l$ being the Fermi wave vector and the mean free path, respectively. In this situation, the exact energies and wave functions in Eq.~\eqref{ns} depend on the disorder configuration, and one needs to look at the average values. 

Before treating the double-tip situation, we calculate the disorder averaged tunneling current for a single tip. The average square modulus of the wavefunction is a constant and, due to the normalization condition, equal to the inverse area ${\cal A}$ of the substrate (assuming the geometry is 2D). Then
\begin{displaymath}
\left\langle n^s (\mathbf{r}, E) \right\rangle = \frac{1}{{\cal A}} \left\langle \sum_k \delta \left( E - E_k \right) \right\rangle = \nu ,
\end{displaymath}
and thus $\langle I (\mathbf{r}, V) \rangle = A \nu eV$. It is position independent and proportional to the voltage.

\subsubsection{Correlations of the tunneling current}

Next, we derive an expression for the current correlations in a double-tip configuration. Our aim is to bring the expression to a form that relates it directly to the observables of the experiment, which are the individual currents and their cumulant, 
\begin{eqnarray} \label{cumulant_def} 
& & J \left( \mathbf{r_1}, \mathbf{r_2}; V_1, V_2 \right) = \left\langle\langle I \left( \mathbf{r_1}, V_1 \right) I \left( \mathbf{r_2}, V_2 \right) \right\rangle\rangle \\
& = & 2 A^2 \int_0^{eV_1} dE_1 \int_0^{eV_2} dE_2 \left\langle\langle n^s (\mathbf{r_1}, E_1) n^s (\mathbf{r_2}, E_2) \right\rangle\rangle \nonumber . 
\end{eqnarray}
The double brackets $\langle\langle \,\, \rangle\rangle$ are defined by $\langle\langle UW \rangle\rangle \equiv \langle UW \rangle - \langle U \rangle\langle W \rangle$. In this expression, the key term to calculate is
 \begin{widetext}
\begin{eqnarray} \label{cumulant_def1}
 & \left\langle\langle n^s (\mathbf{r_1}, E_1) n^s (\mathbf{r_2}, E_2) \right\rangle\rangle = -\nu^2
 + \delta (E_1 - E_2) \left\langle \sum_k \delta (E_k - E_1) \left\vert \psi_k (\mathbf{r_1}) \psi_k (\mathbf{r_2}) \right\vert^2 \right\rangle \\
 & + R(E_1 - E_2) \nonumber \times \left\langle \sum_{k \ne l} \delta(E_1 - E_k) \delta (E_2 - E_l) \left\vert \psi_k (\mathbf{r_1}) \psi_l (\mathbf{r_2}) \right\vert^2 \right\rangle \nonumber,
\end{eqnarray}
\end{widetext}
where $R (\omega)$ is the level-level correlation function. The first term in Eq.~\eqref{cumulant_def1} is just a product of the averages -- it creates a contribution to $J$ which is proportional to $V_1V_2$ and is otherwise position independent and can therefore be ignored. It is the second term that is of interest here and we will refer to it as the same-level correlation term. It describes the correlations of the same state at different points in space and we will estimate it in the following paragraphs. The last term describes the correlations of different states and hence contains the level-level correlation function. We show in Appendix A that in diffusive systems it can also be neglected.

The averages that form the same-level correlation term have been calculated previously in the context of level and wavefunction statistics~\cite{Blanter1997,Mirlin2000} and we only want to sketch the main steps here. For low energies $\vert E_1 - E_2 \vert \ll E_c$, with $E_c \equiv 2\pi \hbar D/L^2$ being the Thouless energy, and for magnetic fields strong enough to break the time-reversal symmetry for a typical electron trajectory (so-called unitary symmetry), we obtain 
\begin{eqnarray} \label{corr1}
\left\langle \sum_k \delta (E_k - E_1) \left\vert \psi_k (\mathbf{r_1}) \psi_k (\mathbf{r_2}) \right\vert^2 \right\rangle = \\
\Delta \nu^2 \left\{ k_d (r) \left[ 1 + \Pi \left(\mathbf{r_1}, \mathbf{r_1} \right) \right] + \Pi \left( \mathbf{r_1}, \mathbf{r_2} \right) \right\}, \nonumber
\end{eqnarray}
and 
\begin{eqnarray} \label{corr2}
\left\langle \sum_{k \ne l} \delta (E_k - E_1) \delta (E_l - E_2) \left\vert \psi_k (\mathbf{r_1}) \psi_l (\mathbf{r_2}) \right\vert^2 \right\rangle = \\
\nu^2 k_d (r) \Pi \left(\mathbf{r_1}, \mathbf{r_1} \right). \nonumber
\end{eqnarray}
Here $r = \vert \mathbf{r_1} - \mathbf{r_2} \vert$, $D$ is the diffusion constant, and $\Delta = (\nu {\cal A})^{-1}$ is the mean level spacing for electrons in the substrate. The short-ranged function $k_d (r) = \exp(-r/l) J_0^2 (k_F r)$ decays at the scale of the Fermi wavelength and, since the tips can not be arranged so closely, does not play a significant role in the correlations. The diffusion propagator $\Pi$ is the solution of the diffusion equation with the corresponding initial and boundary conditions. It can be expressed in terms of the single-electron Green's function as
\begin{equation} \label{propagator}
\Pi \left(\mathbf{r_1}, \mathbf{r_2} \right) = 2 \pi \nu \left\langle G^R \left (\mathbf{r}_1 , \mathbf{r}_2 , \varepsilon \right) G^A \left (\mathbf{r}_2 , \mathbf{r}_1 , \varepsilon \right) \right\rangle \ ,
\end{equation}  
where $G^R$, $G^A$ denote retarded and advanced electron Green's function respectively. Note that the expression does not depend on the energy $\varepsilon$ as soon as it is taken close to the Fermi surface. It is also important that Eqs. (\ref{corr1})-(\ref{propagator}) are general and can be applied to any underlying dynamic of electron motion, not just to the diffusive regime. 

We disregard the terms with $k_d$ and Eq.~\eqref{corr1} becomes
\begin{equation} \label{corr1a}
\left\langle \sum_k \delta (E_k - E_1) \left\vert \psi_k (\mathbf{r_1}) \psi_k (\mathbf{r_2}) \right\vert^2 \right\rangle = \sigma \Delta \nu^2 \Pi \left( \mathbf{r_1}, \mathbf{r_2} ) . \right.
\end{equation}
We will see below that the correlation function of the currents is proportional to the voltage allowing us to directly obtain the diffusion propagator. Note that in Eq.~(\ref{corr1a}) we included both the situation where the external magnetic field is present to break the time-reversal symmetry ($\sigma = 2$), as well as the case where it is absent (orthogonal symmetry, $\sigma=1$).

We now calculate the contribution of the term with $\delta(E_1 - E_2)$ in Eq.~\eqref{cumulant_def1}. The frequency integral is easily calculated, to give
\begin{eqnarray} \label{main_contrib_corr}
& & J_1 \left( \mathbf{r_1}, \mathbf{r_2}; V_1, V_2 \right) \nonumber \\
& = & A^2 \nu^2 \left[ \sigma \Delta \min (eV_1, eV_2) \Pi \left( \mathbf{r_1}, \mathbf{r_2} \right) - e^2 V_1 V_2 \right] . 
\end{eqnarray}
The second term is the product of average currents, and thus the first one (which is much smaller) contains information about the electron states. Note that this term trivially depends on the voltage and on the magnetic field. The presence of the correlations proves that the states spatially extend from $\mathbf{r_1}$ to $\mathbf{r_2}$. The dependence on the tip-to-tip distance given by the diffusion propagator can be probed by repeating the experiment with different double-tip separations.

As we are interested in the autocorrelations of the shared energy levels between the two tips obtained from the current on each of the individual tips, it is not necessary to measure small currents like the transconductance suggested by Niu \textit{et al.}~\cite{Niu1995} and Byers and Flatte~\cite{byers1995probing}. We normalize the correlation function with the individual currents $\langle I_i\rangle=A_i\nu\cdot eV_i$ for $i=1,2$ and setting $V_1=V_2=V$, Eq.~\eqref{main_contrib_corr} then reduces to our final result:
\begin{equation}\label{eq:Jfinal}
\frac{J(\mathbf{r_1},\mathbf{r_2},V,V)}{\langle I_1\rangle \langle I_2\rangle}=\frac{\sigma\Delta}{eV} \Pi(\mathbf{r_1},\mathbf{r_2})-1.
\end{equation} 
This equation directly relates the diffusion propagator to the current correlation normalized by the individual currents. We also note that while the correlations can be long-ranged, one needs nanometer range separation of the tips to measure the diffusion propagator (see numerical estimates below).

\subsubsection{Numerical estimates \& feasibility}

Finally, we present some numerical estimates to show the feasibility of our approach. We consider a Pb nanoisland on which we scan with the two probes. We set the tip separation distance to $r=20$~nm, a spacing that we consider experimentally viable in the near future and that fits well within a 25~nm island. The small volume of islands give rise to a substantial level spacing that we require. Since we prefer an atomically flat surface that fits both tips, we consider a flat island of near monolayer thickness previously obtained~\cite{liu2013growth} and studied by STM~\cite{bose2010observation,kim2011universal,burgess2016closing}. Here we use an already realizable 25~nm island of 3 monolayers thick that has a level spacing of 0.68~meV (for a circular geometry)~\cite{kim2011universal,burgess2016closing}. Our aim is to find an estimate for the signal in Eq.~\eqref{eq:Jfinal} in such a scenario. 

We start by estimating the diffusion coefficient, $D=\frac{1}{2}v_F^2 \tau$, with $\tau$ being the scattering time of the electrons. From the residual resistivity ratio in thin film lead we can determine its low temperature resistivity~\cite{Eichele1981}, and, using Ohm's law, we find the scattering time for Pb
$
\tau={m}/({n e^2 \rho_{4K}})=5.3\times10^{-13}~\textrm{s}.
$
Here, $e$ is the electron charge and we used the low temperature resistivity $\rho_{\textrm{4K}}=9.7\times 10^{-2}~\mu\Omega cm$, the effective electron mass $m=1.9\times m_e$~\cite{Ashcroft} and the total density of valence electrons $n=13.2\times 10^{22}$~cm$^{-3}$. Taken together, this gives us a diffusion coefficient of $D=223$~cm$^2$/s. 

The size of the island and the diffusion constant, as calculated above, result in a Thouless energy
$
E_c={(2\pi \hbar D)}/{\pi L^2}=47~\textrm{meV}\nonumber,
$
and, together with the mean level spacing, the dimensionless conductance
$
g=E_C/\Delta=206.
$
This value satisfies the condition $g \gg 1$ to make the approximation:
$\Pi \left(\mathbf{r_1}, \mathbf{r_2} \right) \approx \frac{1}{g} \ln \frac{L}{r} =1.1 \times 10^{-3}$. 
Including the prefactors, in the absence of a magnetic field and a bias voltage of 1.5~meV, we expect the measured ratio in Eq.~\eqref{eq:Jfinal} to be of the order of $3 \times 10^{-4}$, which can be increased by almost an order of magnitude by moving to monolayer films. We have thus shown that it is in principle possible to measure the diffusion propagator using realistic double-tip parameters. We would also like to note that this is only one example for an experiment using the newly developed smart- and double-tip platforms among many others~\cite{Niu1995,byers1995probing,SettnesPRL,settnesPRB,Ruitenbeek2011,Buttiker1998,Buttiker1999}.

\section{Conclusions} \label{sec:IV}

Motivated by the experimental progress on the realization of double-tip STM's, we present an alternative formalism for probing such electron correlations on the atomic scale. By calculating the current-current correlations between the two tips for disordered metals we have shown that the spacial overlap of the wavefunctions can in fact reach from one tip to the other. In the diffusive limit we also notice that the decay of the signal is logarithmic with increasing tip separation and therefore slower in the ballistic limit~\cite{Niu1995}. To explore the feasibility of the proposed experiment we performed numerical estimates and we show that there is a significant signal to detect. We believe this alternative approach for acquiring electron correlation at the nanoscale may prove interesting as the experiment becomes possible and contribute to reignite interest the largely unexplored possibilities of double-tip STM.   

This project was financially supported by the European Research Council (ERC StG Strong-Q and SpinMelt) and by the Netherlands Organisation for Scientific Research (NWO/OCW), as part of the Frontiers of Nanoscience program, as well as through Vidi grants (680-47-536, 680-47-541).

\bibliography{citations_clean}

\clearpage

\appendix

\section{Level-level correlations} 

This Appendix is to evaluate the contribution of the last term in Eq.~\eqref{cumulant_def1}. We will show that its contribution to $J$ is less important than the first term in Eq.~\eqref{main_contrib_corr}, but that it has non-trivial voltage dependence. Since for a single experiment the distance between the two tips is fixed, we can use this non-trivial voltage dependence to look at correlations between different states in the system.

The level-level correlation function $R(\omega)$ is in the lowest order given by the Wigner-Dyson statistics. The theory of random matrices, from where the Wigner-Dyson statistics originates, discriminates between two situations --- the absence of external magnetic field (Gaussian Orthogonal Ensemble, of GOE, $\sigma = 2$) and the presence of magnetic field (Gaussian Unitary Ensemble, or GUE. $\sigma=1$). In GUE we have
\begin{equation} \label{levelcorr}
R(\omega) = 1 - \left( \frac{\pi\omega}{\Delta} \right)^{-2} \sin^2 \left( \frac{\pi\omega}{\Delta} \right) \ .
\end{equation}

We disregarded the contribution of Eq.~\eqref{corr2} in the main text due to the strong decay of $k_d$ with distance. However, there is a long-range contribution to Eq. \eqref{corr2} which we have not taken into account because it has a higher order in $g^{-1}$ terms in Eq.~\eqref{corr2}. Namely, we have~\cite{Blanter1997,Mirlin2000}
\begin{eqnarray} \label{corr2a}
& & \left\langle \sum_{k \ne l} \delta (E_k - E_1) \delta (E_l - E_2) \left\vert \psi_k (\mathbf{r_1}) \psi_l (\mathbf{r_2}) \right\vert^2 \right\rangle \nonumber \\
& \to & \frac{\sigma}{2}\nu^2 \Pi^2 \left(\mathbf{r_1}, \mathbf{r_2} \right) \ .
\end{eqnarray}

Furthermore, for energies $\omega$ exceeding the Thouless energy $E_c$, $\vert E_1 - E_2 \vert \gg E_c \gg \Delta$, we have 
\begin{widetext}
	\begin{equation} \label{corr2_highen}
	\left\langle \sum_{k \ne l} \delta (E_k - E_1) \delta (E_l - E_2) \left\vert \psi_k (\mathbf{r_1}) \psi_l (\mathbf{r_2}) \right\vert^2 \right\rangle = \frac{\sigma}{2}\nu^2 \mbox{Re} \left[ \Pi^2_{\omega} \left(\mathbf{r_1}, \mathbf{r_2} \right) - \frac{1}{{\cal A}^2} \int d\mathbf{r_2} d\mathbf{r_2} \Pi^2_{\omega} \left(\mathbf{r_1}, \mathbf{r_2} \right) \right] \ 
	\end{equation}
\end{widetext}
and $R_2 = 1$. We again have discarded the short-range terms proportional to $k_d$, assuming that the distance between the tips is much longer than the wavelength. Here, 
\begin{equation} \label{pi-omega}
\Pi_{\omega} \left(\mathbf{r_1}, \mathbf{r_2} \right) = \frac{1}{\pi\nu} \sum_q \frac{\phi_q (\mathbf{r_1}) \phi_q (\mathbf{r_2})}{\hbar Dq^2 - i\omega} \ ,
\end{equation}
where $Dq^2$ and $\phi_q$ are the eigenfunctions and the eigenvalues of the diffusion operator $-D\nabla^2$ with appropriate boundary conditions. Note that at $\omega = 0$, $\Pi_{\omega = 0} (\mathbf{r}_1, \mathbf{r}_2 ) = \Pi (\mathbf{r}_1, \mathbf{r}_2 )$.

To facilitate the calculations, we take $V_1 = V_2 = V$. Since $R(\omega)$ and $\Pi_{\omega}$ are even functions of $\omega$, we can reduce the double integral to a single one using
\begin{equation} \label{doubleint}
\int_0^{eV} dE_1 dE_2 F(E_1 - E_2) = 2 \int_0^{eV} \left( eV - \omega \right) F(\omega) d\omega \ ,
\end{equation}
where $F$ is an arbitrary even function of $\omega$. Due to non-trivial dependences of our functions on $\omega$, we consider different regimes in voltage.

\subsubsection{{Regime 1:} ${eV \ll \Delta}$}

For $eV \ll \Delta$ in GUE we substitute Eq.~\eqref{levelcorr} for $R(\omega)$, calculate
\begin{equation} \label{doubleint1}
\int_0^{eV} \left( eV - \omega \right) R(\omega) d\omega \approx \frac{\pi^2}{36 \Delta^2} (eV)^4 \ ,
\end{equation}
and the contribution to the current correlations from the last term in Eq.~\eqref{cumulant_def1} becomes 
\begin{equation} \label{addterm1}
\delta J \left( \mathbf{r_1}, \mathbf{r_2}; V, V \right) = \frac{\pi^2\nu^2 A^2}{36 \Delta^2} (eV)^4 \Pi^2 \left( \mathbf{r_1}, \mathbf{r_2} \right) \ .
\end{equation}
This differs from Eq.~\eqref{main_contrib_corr} by the factor of $g^{-1} (eV/\Delta)^3 \ll 1$.

For GOE, the level correlation function is cumbersome, but we only need the low-energy behavior, which is $R(\omega) \approx (\pi^2 \vert \omega \vert)/(6 \Delta)$. Calculating the current correlation function, we obtain
\begin{equation} \label{current_GOE}
\delta J \left( \mathbf{r_1}, \mathbf{r_2}; V, V \right) = \frac{\pi^2\nu^2 A^2}{36 \Delta} (eV)^3 \Pi^2 \left( \mathbf{r_1}, \mathbf{r_2} \right) \ .
\end{equation}
It is the same as Eq.~\eqref{addterm1} except for the additional factor $\Delta/eV \gg 1$, making it bigger than Eq.~\eqref{addterm1}. It is still factor $g^{-1} (eV/\Delta)^2 \ll 1$ lower than the contribution of correlations of the same wavefunction.

\subsubsection{{Regime 2:} ${\Delta \ll eV \ll E_c}$}

For $\Delta \ll eV \ll E_c$ we have $R \approx 1$, and, calculating the integral again, we find in both GOE and GUE
\begin{equation} \label{addterm2}
\delta J \left( \mathbf{r_1}, \mathbf{r_2}; V, V \right) = \frac{\pi^2\nu^2 A^2\sigma}{2} (eV)^2 \Pi^2 \left( \mathbf{r_1}, \mathbf{r_2} \right) \ ,
\end{equation}
which is again small compared with Eq.~\eqref{main_contrib_corr} as $eV/E_c \ll 1$.

\subsubsection{{Regime 3:} ${E_c \ll eV \ll \hbar D/r^2}$}

For $eV \gg E_c$, we still have $R = 1$ but now need to use Eq.~\eqref{corr2_highen} to calculate the current-current correlation. To get the results, we now explicitly calculate evaluate $\Pi_{\omega}$ in two dimensions. In \eqref{pi-omega}, we take $\phi_q (\mathbf{r}) = {\cal A}^{-1/2} \exp(i\mathbf{qr})$ and replace the summation over $q$ with integration. Integrating over the angle, we get the Bessel functions, and subsequently integrating over the length of $q$, and we obtain Kelvin functions $\mbox{kei}$ and $\mbox{ker}$,
\begin{eqnarray} \label{pi-omega-expl}
& & \mbox{Re} \ \Pi_{\omega} \left(\mathbf{r_1}, \mathbf{r_2} \right) \\
& = & \frac{1}{4\pi^6 \nu^2 \hbar^2 D^2} \left[ \mbox{kei}^2 \left( \sqrt {\frac{\omega}{\hbar D}} r \right) + \mbox{ker}^2 \left( \sqrt{\frac{\omega}{\hbar D}} r \right) \right] \ . \nonumber 
\end{eqnarray}
In the case of $E_c \ll eV \ll \hbar D/r^2$, we can use the expansion of the Kelvin functions at low arguments, $\mbox{kei}, \mbox{ker} (x) = C, C' - (x/2)^2 \ln (x/2)$, where $C$ and $C'$ are two constants of the order one. We get
\begin{equation} \label{addterm3}
\delta J \left( \mathbf{r_1}, \mathbf{r_2}; V, V \right) \sim \frac{(C+C')\nu^2 A^2\sigma}{2\pi^4 g^2} \frac{(eV)^3}{\hbar D/r^2} \ln \frac{\hbar D/r^2}{eV} \ .
\end{equation}
Comparing this with the first term in $J$, we get
\begin{equation} \label{addterm31}
\delta J/J \sim \frac{1}{\pi^4} \frac{(eV)^2}{E_c \hbar D/r^2} \ln \frac{\hbar D/r^2}{eV} \ , 
\end{equation}
which in principle can become big, but in practice it is unlikely due to the small factor $\pi^{-4}$ in front of this ratio.

\subsubsection{{Regime 4:} ${E_c \gg \hbar D/r^2}$}

In this case, we can replace $(eV - \omega)$ with $eV$ in the integral over $\omega$, and the remaining integral can be calculated exactly. The result is exponentially small ($\exp(-(2eVr^2/\hbar D)^{1/2})$), and does not play any role. 

Note that this regime only makes sense for $r \gg l$ --- then $\hbar D/r^2 \ll 1/\tau$, where $\tau$ is the momentum relaxation time for scattering at impurities. If $eV \gg 1/\tau$, the electron motion at highest energies is not diffusive, and our approach is no longer valid.

\end{document}